\documentclass[aps,prl,preprint,groupedaddress,showkeys]{revtex4-1}

\usepackage{graphicx}
\usepackage{dcolumn}
\usepackage{bm}
\usepackage{braket}
\usepackage{color}

\begin{document}


\title{From the multi-terms urn model to the self-exciting negative binomial distribution \\
and  
\\ Hawkes processes 
}

\author{Masato Hisakado}
\email{hisakadom@yahoo.co.jp} 
\affiliation{
 Nomura Holdings, Inc., Otemachi 2-2-2, Chiyoda-ku, Tokyo 100-8130, Japan.} 
\author{Kodai Hattori}
\email{h21ms111@hirosaki-u.ac.jp}
\author{Shintaro Mori}
\email{shintaro.mori@gmail.com}
\affiliation{
\dag Department of Mathematics and Physics,
Graduate School of Science and Technology, 
Hirosaki University \\
Bunkyo-cho 3, Hirosaki, Aomori 036-8561, Japan}

\date{\today}
\begin{abstract}

This study considers a new multi-term urn process that has a correlation in the same term and temporal correlation.
The objective is to clarify the relationship between the urn model and the Hawkes process.
Correlation in the same term is represented by the P\'{o}lya urn model and the temporal correlation is incorporated by introducing the conditional initial condition.
In the double-scaling limit of this urn process, the self-exciting    negative binomial distribution (SE-NBD) process, which is a marked  Hawkes  process, is obtained. 
In the standard continuous limit, this process becomes the Hawkes process, which has no correlation in the same term.
The difference is the variance of the intensity function in that the phase transition from the steady to the non-steady state can be observed.
The critical point, at which the power law distribution is obtained, is the same for the Hawkes and the urn processes.
These two processes are used to  analyze empirical data of  financial default to estimate the parameters of the model.
For the default portfolio, the results produced by the urn process are superior to those obtained with the Hawkes process and confirm self-excitation.

\hspace{0cm}
\vspace{1cm}
\end{abstract}

\maketitle
\bibliography{basename of .bib file}
\newpage
\section{I. Introduction}
Anomalous diffusion is one of the most interesting topics
in sociophysics and econophysics \cite{galam,galam2,Man}.
Models describing these phenomena have correlations \cite{Bro,W2,G,M,hod,hui,sch}, and
show several types of phase transitions.
In our previous work, we investigated voting models for an information cascade 
\cite{Mori,Hisakado2,Hisakado3,Hisakado35,Hisakado4,Hisakado5,Hisakado6, Mori6}.
This model is a type of urn process that represents the correlations and has two types of phase transitions.
One is the information cascade transition, which is  one of the non-equilibrium phase transitions \cite{Hisakado3}. 
The other is the convergence transition of the
super-normal diffusion that corresponds to an anomalous diffusion \cite{Hod,Hisakado2}.

Financial engineering has led to the development of several products to hedge risks.
These products protect against a subset of the total loss on a credit portfolio in exchange for payments, and
provide valuable insights into the market implications of default dependencies and clustering of defaults \cite{M2010,M2008,Sch}. 
This aspect is important, because the difficulties in managing credit events depend on these correlations.
The Hawkes process was recently used to represent the clustering defaults of time series 
\cite{Haw,kir,qp,err, KZ,KZ1}.
The clustering defaults mean that as the number of events increases, the probability of the events increases. 
This phenomenon corresponds to self-excitation and constitutes a temporal correlation.
From the physical view point 
this process is characterized by a phase transition 
from the steady state to the non-steady state.
It is one kind of the non-equilibrium phase transitions and  the  relation between the  Hawkes process and the branching process  is  shown in \cite{Haw2}. 
In fact the extinction phase  of the branching process corresponds to the steady state in Hawkess process \cite{Hisakado8}.  
Confirmation of the steady state is important for finance and risk management   to hedge risks because it is not possible to manage the non-steady state.



In our previous study,
we discussed the parameter estimation of the urn process, which has a correlation in the same term, and considered a multi-year case with a temporal correlation \cite{Hisakado6, Mori6}.
In this work, we introduce a new extended multi-term urn process and discuss the relationship between the new urn process and the Hawkes process.
In the continuous time limit one can obtain the self-exciting
NBD (SE-NBD) process and Hawkes process as the parameters approach certain
limits.

We study the properties of the phase transitions of these processes.
In fact the simultaneous and temporal effects of the correlation were confirmed by analyzing empirical default data.
The results confirmed that the urn process fits more accurately than the Hawkes process.
The reason is the variance of the  intensity function is nonzero.
In fact  some firms effect many companies and some firms do not effect the companies.
It is one of the  effects of the network, because 
 the network with hubs have the large  variance  of the degree distributions \cite{Hisakado7}.


The remainder of this article is organized as follows.
In Section II, we introduce the multi-term  urn process. We discuss the relationship between the Hawkes process and phase transition.
In Section III we present our study of the phase transition of this model.
In Section IV, we present the power law of the distribution function at the critical point.
In Section V
we discuss the application of the process to empirical
data of historical defaults and confirm its parameters. 
Finally, the conclusions are presented in Section VI.

\section{II. From the multi-term urn process to the Hawkes process }

In this section, we consider a multi-term urn process that has correlations.
In the first term, the urn contains $\theta_0$ red balls and $n_0 -\theta_0$ white balls.
Then, balls are sequentially taken out   from the urn.
For example, a single ball is taken out, after which we return the ball taken out to the urn and  add
$\omega$ balls of the same color to the urn.
Thus, the total number of balls increases by $\omega$ a process, which is a correlation parameter in the same term,
 \cite{Hisakado}.
In fact as we take out  more red balls,  the probability 
 that a red ball  is  taken out increases, because the number of red balls in urn increases. 
 This is the correlation in the same term.
We repeat the process $N$ times in the first term.
Hence, the  number of the total balls is  $n_0+N\omega$ in the end of the first term.
This is nothing but the P\'{o}lya  urn model, which has a beta binomial distribution (BBD).
We summarize the parameters for this urn model in Appendix A.

Next we extend  the process to  the multi-term process. 
We repeat the same process as the first process with the updated parameter $\theta_t$ at the $t+1$-th term. 
In the $t+1$-th term 
the urn contains $\theta_{t}=\theta_0+\sum^{t}_i k_i\hat{d}_{t-i+1}\tilde{\omega}$ red balls, and $n_0 -\theta_t$ white balls. 
Here, the total number of initial balls is $n_0$,
$\hat{d}_i$ denotes the kernel function or  discount factor that represents the
decrease of the effects of  self excitation.
Note that $\theta_t$ depends on the history of number of  the  taken out red balls.
 For the normalization we set $\hat{d}_0=1$.
Here we introduce the variable $X_i$  which 
 is the number of red balls taken out  in the $i$-th term and its value is $X_i=k_i$.
$\tilde{\omega}$ is one of the parameters for temporal correlation, where
$\omega$ denotes the scale parameter for the added red balls.
$\tilde{\omega}$ is the
ratio of the number of balls placed back to the urn to the number of balls 
drawn from the urn, when $\hat{d}_0=1$ and $\hat{d}_i=0, i\ge 1$. In this case  only the previous  term affects the present term. 
In summary we update only the parameter $\theta_t$ and other parameters are fixed in each term.

As the first term we  sequentially take  out  balls from the urn.
After that we return the taken out  ball  to the urn and  $\omega$ balls of the same color are added to the urn.
We repeat the process $N$ times in this $t+1$-th term.
It is the definition of the one term.
After the process of $t+1$ term, we set the initial condition of $t+2$-term and continue the $t+2$-th process.
Each term  is also  the P\'{o}lya urn model.
In this model we use two kinds of correlations:
the first is a correlation in the same term using the parameter $\omega$, and the other is the 
temporal correlation using the parameters $\tilde{\omega}$ which is independent from $\omega$
and $\hat{d}_i$, which is the kernel function.
The temporal correlation decays as a function of time using the parameter $\hat{d}_i$.

In \cite{Hisakado6}, we introduced a similar urn process with a correlation in the same term and temporal correlation.
The difference is the initial condition of the $t+1$-th term.
The number of red balls is the same, but the 
number of white balls is
$n_0-\theta_0+\sum_i^t (N-k_i)\hat{d}_{t-i+1}\tilde \omega$, where $X_i=k_i$, and
the total number of balls in the initial condition of each term is not  constant, $n_0$.

\begin{figure}[h]
\includegraphics[width=110mm]{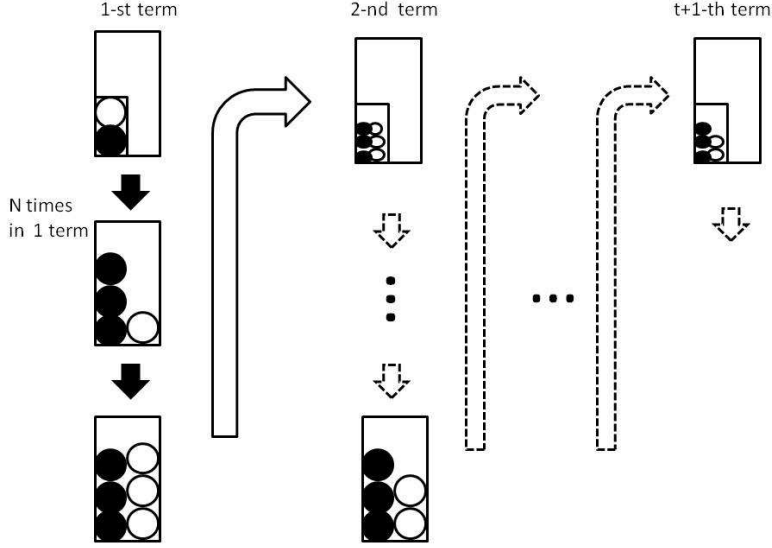}
\caption{ Illustration of the multi-term urn process. The process in each term is a P\'{o}lya urn model.  In each term we repeat the process that we take out  a ball and place back  $\omega$  same collar balls, $N$ times.
In this figure   two balls are taken out in one term, $N=2$.
The process corresponds the black arrow. The conclusion of each term affects the initial condition of the next term as the temporal correlation.  It corresponds to the white arrow. The total number of balls   in the initial condition of each term is $n_0$ and the number of the  red balls in the initial condition of $t$-th term is  $\theta_{t}=\theta_0+\sum^{t}_i k_i\hat{d}_{t-i+1}\tilde{\omega}$. ,
$\hat{d}_i$ denotes the kernel function or  discount factor.
 The number of red balls taken out  in the $i$-th  is $X_i=k_i$. $\tilde{\omega}$ is the scale factor for  the adjustment for the temporal correlation.}
\label{pod}
\end{figure}

When $k_1$ red balls are taken out  in the 1-st term, $X_1=k_1$, the probability in the 1-st term can be calculated as 
\begin{eqnarray}
P(X_1=k_1)
&=&
\frac{N!}{k_1!(N-k_1)!}
[\frac{\theta_0}{n_0}\cdots\frac{\theta_0+(k_1-1)\omega}{n_0+(k_1-1)\omega}]
[\frac{n_0-\theta_0}{n_0+k_1\omega}
\cdots
\frac{n_0-\theta_0+(N-k_1-1)\omega}{n_0+(N-1)\omega}]
\nonumber \\
&=&
\frac{N!}{k_1!(N-k_1)!}
\frac{\Pi_{i=0}^{k_1-1}(\theta_0+i\omega)\Pi_{j=0}^{N-k_1-1}(n_0-\theta_0+j\omega)}
{\Pi_{l=0}^{N-1}(n_0+l\omega)}
\nonumber \\
&=&\frac{N!}{k_1!(N-k_1)!}
\frac{\Pi_{i=0}^{k_1-1}(p+i\rho/(1-\rho))\Pi_{j=0}^{N-k_1-1}(q+j\rho/(1-\rho))}
{\Pi_{l=0}^{N-1}(1+l\rho/(1-\rho))},
\end{eqnarray}
where
$p=\theta_0/n_0$, $q=1-\theta_0/n_0$, and $\omega/n_0=\rho/(1-\rho)$.
This is known as the beta binomial distribution $\mbox{BBD}(\alpha,\beta$,$N$) where 
$p=\alpha/(\alpha+\beta)$ and 
$q=\beta/(\alpha+\beta)$.
Here, $\alpha$ and $\beta$ correspond to the parameters of the beta distribution in the continuous limit of BBD.
Note that $\rho=1/(1+\alpha+\beta)$ plays the role of a correlation in this term \cite{Hisakado}.
In fact  the variance of BBD is 
$Npq+N(N-1)pq\rho$ and the second term corresponds to the
correlation in the same term.
Hence, $\omega$  is  the correlation parameter.

Herein  we consider  the temporal correlation by adjusting the initial conditions of  the  red balls, $\theta_t$ in each term.
The temporal correlation is the decay of the correlation at different times, $\mbox{Cor}(X_t,X_{t+\tau})$  as $\tau$ increases.
The  total number  of  balls in the initial condition  of the each term is $n_0$.
The expected value of the ratio of the taken out  red  balls in $t+1$-th term
is $\theta_t/n_0$, where 
 $\theta_{t}=\theta_0+\sum^{t}_i k_i\hat{d}_{t-i+1}\tilde{\omega}$.
This implies that the events of the previous terms  affect the present events.
Here the  number of the events is the number of the red balls taken out.
As the number of  the events increases, the expected value of the number of events in  the next term increases. 
Hence,  we can  introduce the temporal correlation by 
adjusting  the initial condition of each term. 
When $\theta_t=\theta_0$, there is no temporal correlation, because each term is independent.

Here we set the first  double scaling limit $N/n_0=\Delta$,
$N\rightarrow \infty$ and $n_0\rightarrow \infty$.
We can obtain
\begin{equation}
P(X_1=k_1)\sim \mbox{NBD}(X_1=k_1|K_0, M_0/K_0)=\frac{(K_0+k_1-1)!}{k_1!(K_0-1)!}
(\frac{K_0}{K_0+M_0})^{K_0}
(\frac{M_0}{K_0+M_0})^{k_1},
\end{equation}
where $M_0=\theta_0 N/n_0=\theta_0 \Delta$
and 
$K_0=\theta_0/\omega$.
Here, $\sim$ means that the stochastic variable on the LHS obeys the probability distribution on RHS.
This is a negative binomial distribution (NBD), $\mbox{NBD}(X_1=k_1|K_0,M_0/K_0)$.
Parameter $M_0/K_0=\omega N/n_0=\omega \Delta$ is related to the correlation in this term.
The mean and the variance of NBD is
$M_0$ and $M_0+M_0^2/K_0$, respectively.

The negative binomial distribution
$\mbox{NBD}(X_1=k_1|K_0,M_0/K_0)$ has another face:
\begin{eqnarray}
\mbox{NBD}(X_1=k_1|K_0,M_0/K_0)&=&
\int_0^{\infty}
\mbox{Poisson}(k_1|\lambda) \cdot \mbox{Gamma}(\lambda |K_0,M_0/K_0) d \lambda,
\nonumber \\
&=&
\int_0^{\infty} 
\frac{\lambda^{k_1}e^{-\lambda}}{k_1 !}\dot
\frac{\lambda^{K_0-1}}{\Gamma(K_0)(M_0/K_0)^{K_0}}e^{-\lambda K_0/M_0} d\lambda,
\label{Dec}
\end{eqnarray}
where  $\mbox{Poisson}(k_1|\lambda)$ is the Poisson process $\mbox{Gamma}(\lambda |K_0,M_0/K_0)$ is the gamma distribution.
We show the proof  Eq.(\ref{Dec}) in Appendix B.
From this result  NBD is the Poisson process with the  intensity function $\lambda$ which obeys the gamma distribution. 
In the multi-term model, 
we can decompose the process in the same way.
Hence, $\lambda_t$ is  the intensity function  of the $t$-th term and obeying  the gamma distribution.   
$\mbox{Gamma} (\lambda |K_0,M_0/K_0)$ has an average of $M_0$ and a variance of $M_0^2/K_0$.
This means that the urn process in the 1-st term corresponds to the Poisson process with intensity function $\lambda$, which has a gamma distribution in the double scaling limit.
The intensity function yields the variance of comparing the
Poisson process, which has a constant intensity function.
We refer to this as the NBD process.

\begin{figure}[h]
\begin{center}
\begin{tabular}{c}
\begin{minipage}{0.5\hsize}
\begin{center}
\includegraphics[clip, width=8cm]{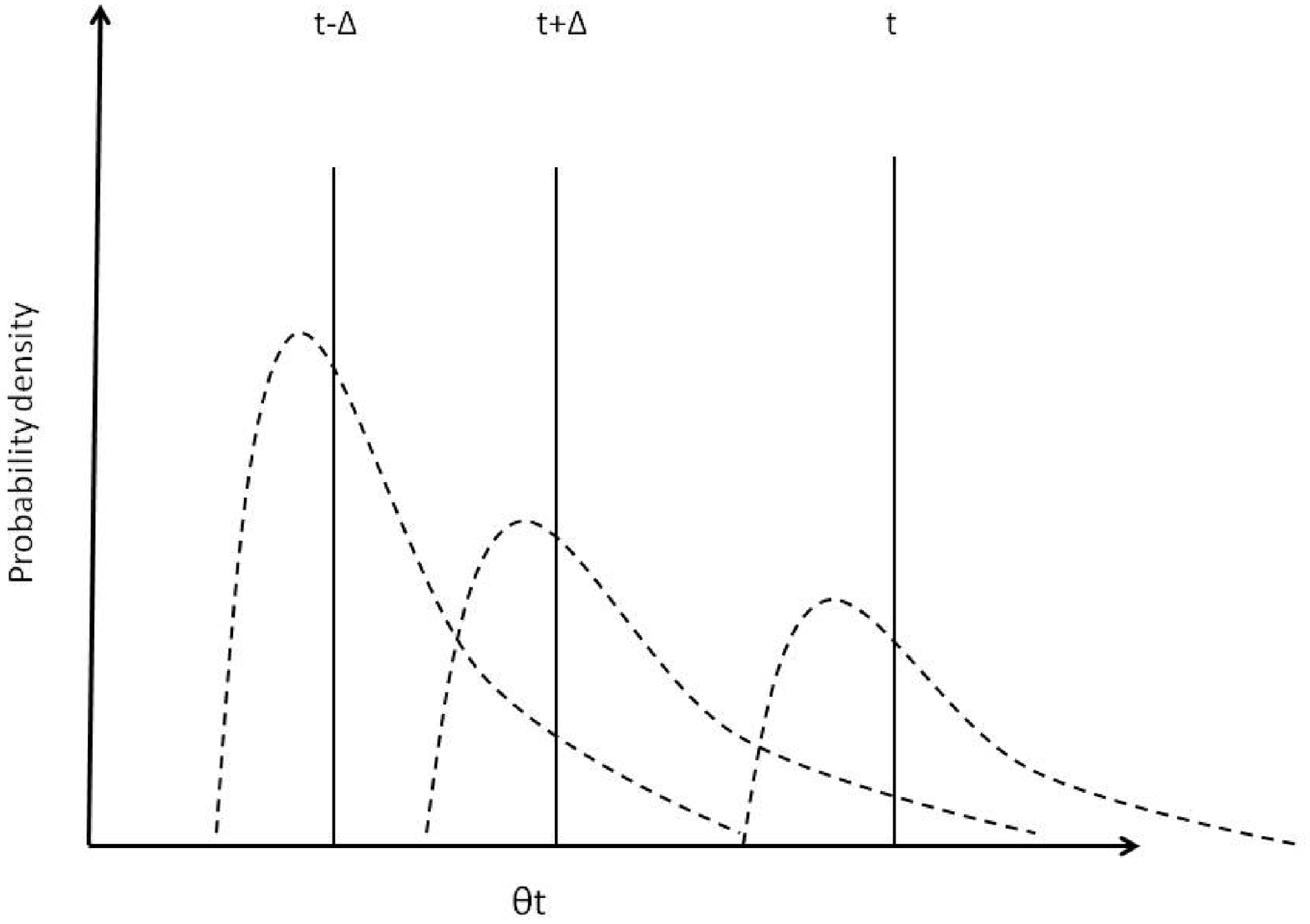}
\hspace{1.6cm} (a)
\end{center}
\end{minipage}
\begin{minipage}{0.5\hsize}
\begin{center}
\includegraphics[clip, width=8cm]{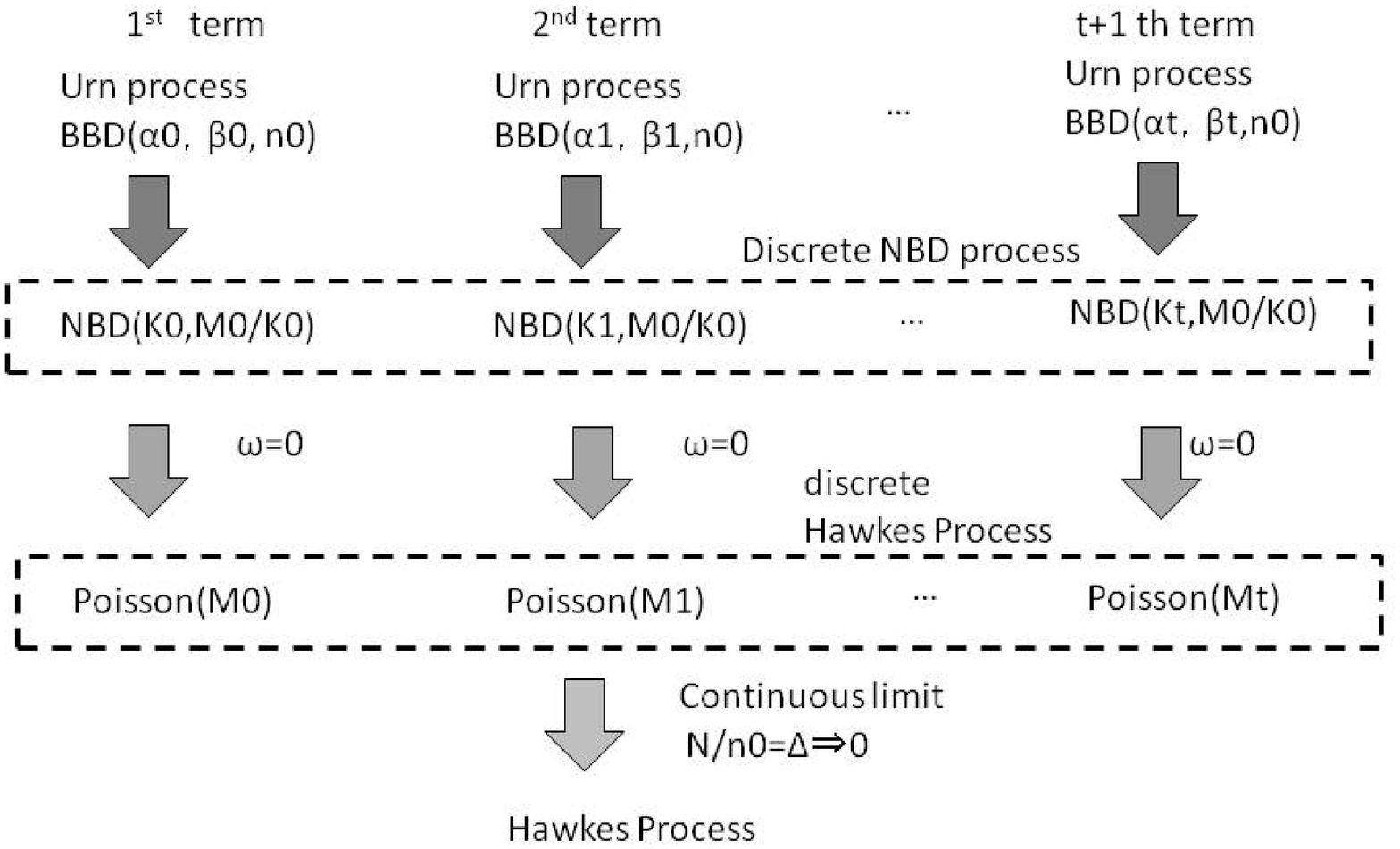}
 \hspace{1.6cm} (b)
\end{center}
\end{minipage}
 \end{tabular}
\caption{Difference between the continuous SE-NBD and Hawkes processes. (a) Intensity function of the continuous SE-NBD process, which obeys the gamma function (dotted line) and Hawkes process, which is the delta function (solid line). It is the  intensity function at $t-\Delta$, $t$, and $\Delta$.
  (b) We can confirm the flow from the BBD to the Hawkes process through the NBD process.}
\label{correlation}
\end{center}
\end{figure}

Next, we define the $t+1$-th term with the temporal correlation \cite{Hisakado6} using the conditional probability.
We define the conditional probability of $t+1$-th term,
\begin{equation}
    P(X_{t+1}=k_{t+1}|X_0=k_0, \cdots, X_{t}=k_{t})
    =\mbox{NBD}(X_{t+1}=k_{t+1}|K_t, M_t/K_t),
    \label{cp}
\end{equation}
where $k_i$ is the history of the number of red balls taken out.
The conditional probability is defined by updating parameters $K_t$ and $M_t$.
The only difference between the 1-st term and the $t+1$-th term is the number of white balls in the initial condition of the term.
The other conditions are the same as those in the 1-st term.
We can obtain the parameters at the $t+1$-th term for the intensity function:
\begin{eqnarray}
M_{t}&=&\theta_t N/n_0=\frac{\theta_0+\sum_i^{t}X_i \hat{d}_{t+1-i} \tilde{\omega}}{n_0}N
=(\theta_0+\sum_i^{t}X_i \hat{d}_{t+1-i} \tilde{\omega})\Delta
\nonumber \\
&=&M_0+M_0/L_0\sum_i^{t}X_i \hat{d}_{t+1-i},
\label{m}
\end{eqnarray}
where $M_0=\theta_0 N/n_0=\theta_0 \Delta$, $K_0=\theta_0/\omega$, $L_0=\theta_0/\tilde{\omega}$
 and $\omega N/n_0=\omega \Delta=M_0/K_0$.
The other parameters are obtained as follows:
\begin{equation}
K_{t}=\theta_t /\omega=\frac{\theta_0+\sum_i^{t}X_i\hat{d}_{t+1-i} \tilde{\omega}}{\omega}= 
K_0+K_0/L_0\sum_i^{t}k_i\hat{d}_{t+1-i},
\end{equation}
and 
\begin{equation}
M_{t}/K_t=\omega N/n_0=\omega \Delta=M_0/K_0.
\end{equation}
We refer to this  process  as the discrete self-exciting negative binomial distribution (SE-NBD) process.
The self-exciting is introduced by the conditional probability,
Eq. (\ref{cp}).
Note that for all the process parameters, $M_t/K_t$ is a constant $M_0/K_0$.
This signifies that the correlation in the same term 
 does not depend on the term $t$. 
By this condition the process has the reproductive property of NBD.
The mean of the intensity function is
$M_t$ and the variance is $M_t^2/K_t$.
As $\omega$ increases, $K_t$ decreases, and the variance
of the intensity function increases.
In this mean the correlation in the same term affects the
variance of the intensity function.
In
 the limit $K_0\rightarrow \infty(\omega\rightarrow 0)$ with $\Delta$ fixed, 
 the intensity function has a vanishing variance and thus its distribution converges to the delta function.
  This is known as the discrete Hawkes process \cite{kir}.
 The intensity function is illustrated in Fig.\ref{correlation} (a).
$L_0$ acts as a parameter for the temporal correlation as $\tilde{\omega}$.

In summary, we
obtained that the discrete SE-NBD process
$X_{t}$ obeys NBD for $M_t$
from the urn process,
\begin{equation}
X_{t+1}\sim \mbox{NBD}
\left(K_{t},M_0/K_0 \right),t\ge 0, 
\end{equation}
where 
\begin{equation}
M_t=M_0+M_0/L_0\sum_{s=1}^{t}X_s \hat{d}_{t+1-s},t\ge 1,  
\end{equation}
and 
\begin{equation}
K_t=K_0+K_0/L_0\sum_{s=1}^{t}X_s \hat{d}_{t+1-s},t\ge 1.  
\end{equation}
In the limit $K_0 \rightarrow \infty (\omega=0)$, the discrete Hawkes process,
$X_{t}$, is shown to obey the Poisson process for $M_t$
from the urn process,
\begin{equation}
X_{t+1}\sim \mbox{Poisson}
\left(M_{t} \right),t\ge 0, 
\end{equation}
where 
\begin{equation}
M_t=M_0+M_0/L_0\sum_{s=1}^{t}X_s \hat{d}_{t+1-s},t\ge 1.  
\end{equation}
In the limit $L_0 \rightarrow \infty (\tilde{\omega}=0)$, the process becomes an NBD process, which does not have self-excitation capabilities. 
$\Delta$ corresponds to the number of balls taken out  in a term, and represents 
the interval between the terms.
Here we introduce the counting process, $\tilde{X}_t=\sum_{i} X_i$.
We set the  second double scaling limit $\Delta=N/n_0\rightarrow 0$, $\omega\rightarrow \infty$ with $\omega\Delta=\omega'$, as the continuous limit of the process $\tilde{X}_t$.
Note that in this limit $\omega\Delta=\omega'$ is constant and 
the process has the reproductive property as a discrete SE-NBD process. 
We can obtain the mean of the intensity function at $t$, $\lambda_t$
\begin{equation}
E(\lambda_t|F_t)=\lim_{\Delta\rightarrow 0}\frac{E[\tilde{X}_{t+\Delta}-\tilde{X}_t|F_t]}{\Delta}=\lim_{\Delta\rightarrow 0}\frac{M_t}{\Delta}=(\theta_0+\tilde{\omega}\sum_{i<t} X_i\hat{d}_{t-i} ),
\label{av}
\end{equation}
where $F_t$ is the history of the number of red balls and $X_1=k_1, \cdots, X_t=k_t$ and 
it corresponds to the Hazard function \cite{sch}.
The variance of the intensity of the distribution at time $t$ is 
\begin{equation}
Var(\lambda_t|F_t)=\lim_{\Delta\rightarrow 0}\frac{M_t^2/K_t}{\Delta}=\omega'   ( \theta_0+\tilde{\omega}\sum_{i<t} \hat{d}_{t-i} X_i).
\label{ba}
\end{equation}

In the continuous SE-NBD process, the intensity function exhibits a gamma distribution as a discrete SE-NBD process.
We can confirm that the intensity function becomes a delta function in the limit $\omega\rightarrow 0$, which corresponds to the continuous Hawkes process.
In summary, we can obtain in the continuous limit, 
\begin{equation}
\tilde{X}_{t+\Delta}-\tilde{X}_{t}\sim \mbox{NBD}
\left(\theta_t\Delta/\omega',\omega'\right),t\ge 0, 
\label{av2}
\end{equation}
where 
\begin{equation}
\theta_t=\theta_0+\tilde{\omega} \sum_{s<t} X_s\hat{d}_{t-s},t\ge 0.  \label{SE}
\end{equation}

In the limit $\omega'\rightarrow 0$, the continuous SE-NBD process becomes the Hawkes process.
\begin{equation}
\tilde{X}_{t+\Delta}-\tilde{X}_{t}\sim \mbox{Poisson}
\left(\theta_t\Delta \right),t\ge 0, 
\label{av3}
\end{equation}
where 
\begin{equation}
\theta_t=\theta_0+\tilde{\omega} \sum_{s<t} X_s\hat{d}_{t-s},t\ge 0.  \label{HP}
\end{equation}
We show the path form the discrete Hawkes process to the Hawkes process in Appendix C. 
Finally, we show the path from the urn process to SE-NBD process and the Hawkes process in
Fig. \ref{correlation} (b).

\section{III. Phase transition of the new urn process}

In this section we consider the phase transition of the SE-NBD process.
Here  we apply the mean field approximation  that   we set $\bar{v}=X_t$.
We consider the increase of the process in $\Delta$,
\begin{equation}
    E[\tilde{X}_{t+\Delta}-\tilde{X}_t]=E[\lambda_t|F_t]
    =[\theta_0+\bar{v}\tilde{\omega}\sum_i\hat{d}_i]\Delta,
\label{mfe}
\end{equation}
 where we use Eq.(\ref{SE}).
We set the average, $\bar{v}$, of the intensity function
 and LHS of Eq.(\ref{mfe}) is $\bar{v}\Delta$.
Then we  obtain,  
\begin{equation}
\bar{v}=\theta_0/(1-\tilde{\omega} \hat{T}),
\end{equation}
 where $\hat{T}=\sum^{\infty} \hat{d}_{i}$.

In the limit $\tilde{\omega}\rightarrow 0 $, the temporal correlation is zero and the process is  the  NBD process,
where  the phase transition disappears.

The SE-NBD includes the Hawkes process as a branching process.
The branching ratio is
\begin{equation}
\nu=  \tilde{\omega}\hat{T},
\end{equation}
and the condition for the steady state is
\begin{equation}
\nu= \tilde{\omega}\hat{T} <1.
\end{equation}
The phase transition between the steady state and  non-steady state occurs at $\nu=1$, which is the critical point.
 The transition point is  the same as that in the Hawkes process \cite{KZ,KZ1}.
 
The parameter $\nu$ is  termed the effective reproduction number
with regard to an infectious disease.
It is the number of patients infected by one patient 
in the infection model.
If the effective reproduction number is above 1, the
number of patients increases to infinity  indicating  the non-steady state.

In the SE-NBD process, the distribution of   the intensity function  has   variance.
By contrast, in the Hawkes process, the intensity function is a delta function.
The variance of the intensity function is the origin of the 
variance of the branching ratios.
This  means that  the SE-NBD process has the variance of  branching ratios, because the intensity function has the variance. 
In fact, \cite{nishi} demonstrated  that the effective reproduction number depends on the COVID-19 environment.
The mixture of branching ratios affects not only the expected value of the
intensity function but also
the variance of the intensity function.
Hence, the SE-NBD process, which has  gamma distribution
as the intensity function, may be useful.
We confirm this in Section V.

We consider the exponential and power decay cases
as the kernel function.
These  correspond to  short  and long memories \cite{Long} as the kernel function.
When we consider the exponential decay case
$\hat{d}=e^{-\beta t}$, 
the condition for the steady state is $M_0/L_0<\beta$.
When we consider the power decay case
$\hat{d}=1/(1+t)^{\gamma}$, 
the condition for the steady state is $M_0/L_0<\gamma-1$. 
In Section V we use the exponential kernel for the 
empirical default data.

\section{IV. Power-law distribution at critical point}

We start with a discrete SE-NBD process $\{X_{t}\},t=1,\cdots$. Here, $X_t\in \{0,1,\cdots\}$ represents the size of an event at time $t$.
This is the process we obtained from the urn process.
This event corresponds to the taken out  of the red ball from the urn.
$X_{t}$ obeys NBD for $M_t$.
\begin{eqnarray}
X_{t+1}&\sim& \mbox{NBD}
\left(\frac{M_{t}}{\omega}=K_t,\omega \Delta\right),t\ge 0, \nonumber \\
M_t&=&M_0+n\sum_{i=1}^{t}X_s h(t-i),t\ge 1,  \nonumber
\end{eqnarray}
where $n=M_0/(1-r)L_0$.
We adopt the exponential decay kernel function, $h(t)=(1-r)\hat{d}=(1-r)r^t,0\le r<1$.
In addition, we replace the normalization factor $(1-r)$ of $h(t)$
with $1/\tau$ to ensure that $\int_{0}^{\infty}h(t)dt=\frac{1}{\tau}\int_{0}^{\infty}e^{-t/\tau}dt=1$.

The stochastic process $\{X_t\},t=1,\cdots$ is non-Markovian. We focus on the
time evolution of the intensity function $M_{t}$, which
satisfies the following recursive equation.
\[
M_{t+1}=r(M_{t}-M_0)+M_0+nh(0)X_{t+1}.
\]
Here, we use the relationship
$\sum_{i=1}^{t+1}X_{i}h(t+1-i)=X_{t+1}h(0)+r\sum_{i=1}^{t}X_{s}h(t-i)$.
The stochastic difference equation for the excess intensity
$\hat{z}_{t}\equiv M_t-M_0$ is
\begin{equation}
 \hat{z}_{t+1}-\hat{z}_{t}=(r-1)\hat{z}_{t}+nh(0)X_{t+1}.
\end{equation}  

We take the continuous time limit as in Section II. We divided the unit time interval
by the infinitesimal time intervals with width $dt=\Delta$.
The decreasing factor $r^t$ during the
interval $dt$ is replaced with $r^{dt}=e^{-dt/\tau}\simeq 1-dt/\tau+o(dt/\tau)$.

$X_{t+1}$ is the noise for time interval $[t,t+1]$, and it is necessary to
prepare the noise for the infinitesimal interval $[t,t+dt)$.
For the purpose, we use the reproductive property of NBD.
If $X_{t+1}$, the noise for the interval $[t,t+1)$ obeys $\mbox{NBD}(\theta_t/\omega,\omega)$.
The noise for $[t,t+dt)$  is denoted by  $d\hat{\xi}^{NB}_{(\theta_t/\omega',\omega')}(t)$.
Here, the double scaling limit is applied to 
change the parameter from $\omega$ to $\omega'$.
As Eq.(\ref{av2}) 
the stochastic difference equation (SDE) then becomes
\begin{equation}
d\hat{z}_{t}=\hat{z}_{t+dt}-\hat{z}_{t}=-\frac{1}{\tau}\hat{z}_{t}dt+\frac{n}{\tau}
d\hat{\xi}^{NBD}_{(\frac{\theta_0+\hat{z}_{t}}{\omega'},\omega')}(t). \label{eq:SDE}
\end{equation}
The state-dependent NBD noise $\hat{\xi}^{NBD}_{\theta_t/\omega,\omega}$ defines the noise
for the infinitesimal time interval $[t,t+dt)$ with
the following probabilistic rules:
\[
d\hat{\xi}^{NBD}_{(\theta_t/\omega',\omega')}(t)\equiv
\hat{\xi}^{NBD}_{(\theta_t/\omega',\omega')}(t+dt)-\hat{\xi}^{NBD}_{(\theta_t/\omega',\omega')}(t)\sim
\mbox{NBD}(\theta_t dt/\omega',\omega').
\]
The characteristic function for $d\hat{\xi}^{NBD}_{(\theta_t/\omega',\omega')}$
is $\phi(s)=(\frac{1}{1+\omega'-\omega' e^{is}})^{\theta_t dt/\omega'}$. The infinite divisibility is
  clear from the functional form. In addition,
  $d\hat{\xi}^{NBD}_{(\theta_t/\omega',\omega')}$ 
  can be written as the superposition of the Poisson noise
  with the state-dependent random intensity $\lambda_t$,
\[
d\xi^{NBD}_{(\theta_t/\omega',\omega')}\sim \mbox{Poisson}(\lambda_t dt)
,\lambda_t\sim \mbox{Gamma}(\theta_t/\omega',\omega').
\]
 as  we discussed in section II.
When we denote the timing and size of the $i$-th event as $t_i$ and $k_{i}$, 
and the number of events before $t$ as $\hat{N}(t)$, we can rewrite
the state-dependent NBD noise $\hat{\xi}^{NBD}_{(\frac{\theta_0+\hat{z}_{t}}{\omega' },\omega')}(t)$ as
\[
\hat{\xi}^{NBD}_{(\frac{\theta_0+\hat{z}_{t}}{\omega' },\omega')}(t)
=\sum_{i=1}^{\hat{N}(t)}k_{i}\delta(t-t_i).
\]
The probability of the occurrence and non-occurrence of an event of size $k$
during time interval $dt$ is given as,
\begin{equation}
P\left(d\hat{\xi}^{NBD}_{(\frac{\theta_0+\hat{z}_{t}}{\omega' },\omega')}=k\right)=
\left\{
\begin{array}{cc}
1-\frac{\hat{z}_{t}+\theta_0}{\omega' }\ln(\omega' +1)dt  & k=0 \\
\frac{1}{k}
\left(\frac{\hat{z}_t+\theta_0}{\omega' }\right)
\left(\frac{\omega'}{\omega' +1}\right)^kdt &  k\ge 1. 
\end{array}
\right.
\label{NBD}
\end{equation}
In the limit $\omega \to 0$, the probabilities becomes
\[
\lim_{\omega  \to 0}
P\left(d\hat{\xi}^{NBD}_{(\frac{\theta_0+\hat{z}_{t}}{\omega' },\omega')}=k\right)=
\left\{
\begin{array}{cc}
1-(\hat{z}_{t}+\theta_0)dt  & k=0 \\
(\hat{z}_{t}+\theta_0)dt  & k=1 \\
0   &  k\ge 2. 
\end{array}
\right.
\]
$k_i$ is restricted
to be one or zero, and the state dependent noise becomes the Poisson noise.
We discuss the relation  between the SE-NBD and the marked   Hawkes processes in Appendix D.

The SDE (\ref{eq:SDE}) is interpreted as
\[
\hat{z}(t+dt)-\hat{z}(t)=
\left\{
\begin{array}{cc}
-\frac{1}{\tau}\hat{z}_t & \mbox{Prob.}=1-\frac{\hat{z}_t+\theta_0}{\omega' }\ln (\omega'  +1)dt \\
\frac{nk}{\tau}  & \mbox{Prob.}=\frac{1}{k}\left(\frac{\hat{z}_t+\theta_0}{\omega' }\right)\left(\frac{\omega'}{\omega'+1}\right)^kdt, k=1,\cdots.
\end{array}  
\right.
\]

The same procedure is adopted
to derive the master equation for the probability density function (PDF) of $\hat{z}_{t}$ in \cite{KZ,KZ1}.
This yields 
\begin{equation}
\frac{\partial}{\partial t}P_{t}(z)=\frac{1}{\tau}\frac{\partial}{\partial z}
z P_{t}(z)+\sum_{k=1}^{\infty}\frac{1}{\omega'  k}\left(\frac{\omega' }{\omega'  +1}\right)^k
\left\{(\theta_0+z-\frac{nk}{\tau})P_{t}(z-\frac{nk}{\tau})-(\theta_0+z)P_{t}(z)\right\}.
\end{equation}  
The Laplacian representation of the steady state is denoted as
$P_{SS}(z)$ as $\tilde{P}_{SS}(s)$.
\[
\tilde{P}_{SS}(s)\equiv \int_{0}^{\infty}P_{SS}(z)e^{-sz}dz.
\]
The master equation for $\tilde{P}_{SS}(s)$ is
\begin{eqnarray}
\left[\sum_{k=1}^{\infty}\frac{1}{\omega'  k}\left(\frac{\omega' }{\omega' +1}\right)^k
  \left(e^{-\frac{nk}{\tau}s}-1\right)+\frac{s}{\tau} \right]\frac{d}{ds}\tilde{P}_{SS}(s)=\sum_{k=1}^{n}\frac{1}{\omega k'}\left(\frac{\omega'}{\omega'+1}\right)^k
\theta_0 \left(e^{-\frac{nk}{\tau}s}-1\right)\tilde{P}_{SS}(s). \nonumber \\
\end{eqnarray}
Thus, 
\[
\frac{d}{ds}\ln \tilde{P}_{SS}(s)=\theta_0-\frac{\theta_0
  s/\tau}{\sum_{k=1}^{\infty}\frac{1}{\omega'  k}
  \left(\frac{\omega' }{\omega' +1}\right)^k(e^{-\frac{nk}{\tau}s}-1)+\frac{s}{\tau}}.
\]
We integrate the equation with the initial condition $\tilde{P}_{SS}(0)=1$.
\[
\ln \tilde{P}_{SS}(s)=\theta_0 s-{\displaystyle \int_{0}^{s}\frac{\theta_0
  s'/\tau}{\sum_{k=1}^{\infty}\frac{1}{\omega'  k}
  \left(\frac{\omega' }{\omega' +1}\right)^k(e^{-\frac{nk}{\tau}s'}-1)+\frac{s'}{\tau}}}ds'.
\]
Here, the large $z$ behavior of $P_{SS}(z)$ near the critical
point $n=1$ is of interest, and we study the
integral at $s\simeq 0$ and $n=1-\epsilon,\epsilon<<1$.
We expand $e^{\frac{nk}{\tau}s}=1-\frac{nks}{\tau}+\frac{1}{2}(\frac{nks}{\tau})^2+o(s^2)$
and calculate the summation over $k$ in the denominator of the integral.
Therefore, we have
\[
\ln \tilde{P}_{SS}(s)\simeq \theta_0 s-\int_{0}^{s}\frac{\frac{2\theta_0\tau}{\omega' +1}}{\frac{2\tau}{\omega' +1}\epsilon+s'}ds'.
\]
Near the critical point, the excess intensity distribution shows
power-law behavior with a non-universal exponent,
up to an exponential truncation:
\begin{equation}
  P_{SS}(z)\propto z^{-1+2\frac{\theta_0 \tau}{\omega' +1}}e^{-\frac{2\tau\epsilon}{\omega' +1}z}
\label{eq:ss}.
\end{equation}
The power-law exponent of the PDF of the excess intensity is
$1-\frac{2\theta_0 \tau}{\omega'+1}$, and depends on $\omega'$, which is the correlation simultaneously.
In the limit $\omega'\to 0$, the result coincides with that in \cite{KZ,KZ1}. The power-law exponent increases with $\omega'$ and
it converges to 1 in the limit $\omega'\to \infty$. 
$\omega' \rightarrow 0$ is the no  correlation  in the same time and the intensity function becomes the delta function. 
$\omega'\rightarrow \infty$ is the strong correlation  limit  in the same time and the intensity function has  the large variance.  
The correlation
in the same term alters the critical behavior. In addition,
the length scale 
of the exponential decay for the off-critical case is
$(\omega'+1)/(2\tau\epsilon)$, which is also an increasing function of $\omega'$.

\section{V. Parameter estimation using default data }

We use empirical data pertaining to financial default to estimate the parameters.
First, the S\&P default data from 1981 to 2020 are used. 
A speculative grade (SG) rating represents ratings below BBB-(Baa3), whereas an investment grade (IG) rating indicates those above BBB-(Baa3).
We  also use Moody's default data from 1920 to 2020, a period of 100 years that includes the Great Depression of 1929 and the Great Recession of 2008.

We estimate the parameters using Bayes’ formula
\begin{eqnarray}
P(K_0, M_0,L_0,\beta|k_0, k_1, \cdots, k_T)
&=&\frac{P(r_T|K_0,M_0,L_0, \beta,))}{P(k_T)}
\cdots
\frac{P(k_0|K_0,M_0,L_0,\beta)}{P(k_0)}
\nonumber \\
&\times&f(K_0,M_0,L_0,\beta),
\label{MAP}
\end{eqnarray}
where $f(K_0,M_0,L_0,\beta)$ is a prior distribution \cite{Hisakado6}, for which we used a uniform distribution.
We apply  the maximum a posteriori (MAP) estimation expressed by Eq.(\ref{MAP}).
The use of the NBD to determine the distribution $P$ would be the parameter estimation for the discrete SE-NBD process introduced in Section II.
The use of the Poisson distribution instead of the NBD would be the parameter estimation for the discrete Hawkes process.
We present the outcome of the optimization using the discrete SE-NBD, discrete Hawkes, and NBD processes in Tables \ref{game13}, \ref{game23}, and \ref{game14}.
NBD was employed to confirm whether self-excitation existed.
For the Hawkes process, $K_0\rightarrow \infty (\omega=0)$, and for the NBD process, $L_0\rightarrow \infty(\tilde{\omega}=0)$, as discussed in Section II.

When $K_0$ is large, it is nearly a Hawkes process.
When $L_0$ is large, it is nearly an NBD process, which has no self-excitation. 
The estimated $K_0$ is small for the SE-NBD process and, particularly, $K_0$ is small for IG.  
 As in Fig. \ref{game14}, we can obtain a much better  AIC for the SE-NBD process.
 This implies an intensity function of which the variance is not a delta function, as in the Hawkes process.
 In fact, certain defaulted obligors affect other obligors, whereas this does not occur in the case of other obligors.
 The former case corresponds to the situation of chain bankruptcy and may be  considered to depend on network effects.
 An obligor who is connected to many obligors 
 affects many other obligors.
 In fact, the AIC for the SE-NBD process was smaller than that for the NBD process.
 Hence, we can confirm the existence of self-excitation in this historical credit dataset.

\begin{table}[tbh]
\caption{MAP estimation of the parameters for the discrete SE-NBD and discrete Hawkes processes}
\begin{center}
\begin{tabular}{|c|l|lllll|cccc|r|}
\multicolumn{2}{c}{}\\ \hline
&  & SE-NBD  & &&&  & Hawkes&  & &\\
No.& Model & $K_0$&$L_0$& $M_0/K_0$ & $\beta$&$\bar{v}$ & $M_0$ & $L_0$ &$\beta$ & $\bar{v}$  
\\ \hline \hline
1&Moody's 1920-2020& 0.28&6.17& 18.89 & 2.94&
58.35& 3.4& 3.55&15.98 &86.85\\ \hline
2&S\&P 1981-2020& 1.06&27.80& 18.95 & 16.08&71.22& 13.3 & 15.76& 18.40& 83.65 \\ \hline
3& Moody's 1981- 2020&1.03 &32.12 & 22.55&15.97&82.79& 17.2 & 21.09 &16.69 &91.72 \\ \hline
4&S\&P 1990-2020& 1.51&62.52 & 23.04 & 16.23&78.64& 29.2 & 45.37& 13.01& 81.82  \\ \hline
5&Moody's 1990-2020& 1.58&86.55 & 27.92& 14.40&90.00& 40.6& 73.03&19.19 &91.38 \\ \hline
6&Moody's 1920-2020 SG &0.29 &5.91 & 17.81 & 3.03&56.04& 2.9 & 3.00 & 13.57&105.32 \\ \hline
7&S\&P 1981-2020  SG& 1.05&25.66 & 17.90 & 16.22&69.90& 12.0 & 13.93 & 17.57&85.29  \\ \hline
8&Moody's 1981-2020 SG&1.02 & 30.57 & 21.65 & 15.99&80.65 &15.4 & 18.53& 15.71&92.38\\ \hline
9&S\&P 1990-2020  SG& 1.54&60.05 & 21.82 & 16.02&76.39& 28.2 & 43.74 & 18.97&79.59 \\ \hline
10&Moody's 1990-2020  SG&1.62 &86.79 & 26.76 & 15.44&87.05& 39.6 & 71.53 & 14.57&88.57 \\ \hline
11&Moody's 1920-2020 IG & 0.13&1.10 & 4.06 & 0.99&2.13& 1.14 & 0.40 &0.98 &1.24 \\ \hline
12&S\&P  1981-2020 IG& 0.39&2.87 & 3.06 & 15.32&2.05 &1.2 & 2.67 & 14.68 &2.05\\ \hline
13&Moody's 1981-2020 IG&0.28&6.07 & 5.37 & 1.26&2.36 &1.7 & 6.02 & 14.63 &2.34\\ \hline
14&S\&P 1990-2020  IG&0.33 &2.73& 3.75 & 16.80&2.32& 1.2 & 2.65 & 17.46&2.32\\ \hline
15&Moody's 1990-2020  IG& 0.28&4.13 & 5.80 & 13.27&2.69& 1.8 &5.58  &16.26 & 2.68\\ \hline
\end{tabular}
\label{game13}
\end{center}
\end{table}

\begin{table}[tbh]
\caption{MAP estimation of the parameters for the NBD process}
\begin{center}
\begin{tabular}{|c|l|lll|}
\multicolumn{2}{c}{}\\ \hline
&  & & NBD& \\
No.& Model & $K_0$& $M_0/K_0$ & $\bar{v}$   
\\ \hline \hline
1&Moody's 1920-2020& 0.47&80.64& 37.86 \\ \hline
2&S\&P 1981-2020& 1.54&38.61& 59.62 \\ \hline
3& Moody's 1981- 2020&1.52 &45.76& 69.78\\ \hline
4&S\&P 1990-2020& 2.07&34.37 & 71.29 \\ \hline
5&Moody's 1990-2020& 2.14&38.86 & 83.23\\ \hline
6&Moody's 1920-2020 SG &0.47 &76.28 & 35.81 \\ \hline
7&S\&P 1981-2020  SG& 1.55&37.14& 57.57 \\ \hline
8&Moody's 1981-2020 SG&1.53 & 44.00 & 67.45 \\ \hline
9&S\&P 1990-2020  SG& 2.12&32.46 & 68.97\\ \hline
10&Moody's 1990-2020  SG&2.20 &36.65 & 80.58 \\ \hline
11&Moody's 1920-2020 IG & 0.29&7.18 & 2.05 \\ \hline
12&S\&P  1981-2020 IG& 0.46&4.42 & 2.05 \\ \hline
13&Moody's 1981-2020 IG&0.37&6.30& 2.33 \\ \hline
14&S\&P 1990-2020  IG&0.41 &5.63& 2.32 \\ \hline
15&Moody's 1990-2020  IG& 0.36&7.32 & 2.65 \\ \hline
\end{tabular}
\label{game23}
\end{center}
\end{table}

\begin{table}[tbh]
\caption{AIC for the discrete SE-NBD, discrete Hawkes, and NBD processes }
\begin{center}
\begin{tabular}{|l|l|c|c|c|r|}
\multicolumn{4}{c}{}\\ \hline
&   &SE-NBD process&   Hawkes process &   NBD process \\
No.& Model & 
 AIC & AIC &AIC\\
 \hline \hline
1&Moody's 1920-2020& 791.9& 2193.1 &904.0\\ \hline
2& S\&P  1981- 2020& 386.7 & 1010.6 &407.9\\ \hline
3& Moody's 1981-2020& 399.3 & 1186.1 &420.6\\ \hline
4&S\&P 1990-2020& 316.3 & 923.0 &323.5\\ \hline
5&Moody's 1990-2020& 327.1 & 1098.6 &332.4\\ \hline
6&Moody's 1920-2020 SG & 781.5 & 2060.1 &893.0 \\ \hline
7&S\&P 1981-2020 SG&383.2 & 975.8 &405.1\\ \hline
8&Moody's 1981-2020  SG& 396.5& 1140.0&417.8 \\ \hline
9&S\&P 1990-2020  SG& 313.9& 894.0&321.0 \\ \hline
10&Moody's 1990-2020  SG& 325.1& 1062.4 &329.9 \\ \hline
11&Moody's 1920-2020 IG & 321.7 & 490.1  &360.3 \\ \hline
12&S\&P 1981-2020 IG&150.0 & 197.6 &153.2 \\ \hline
13&Moody's  1981-2020 IG&156.8& 257.1& 156.8\\ \hline
14&S\&P 1990-2020  IG& 121.3& 168.3 & 124.2 \\ \hline
15&Moody's 1990-2020  IG& 127.4 & 219.7 &127.9  \\ \hline
\end{tabular}
\label{game14}
\end{center}
\end{table}

\section{VI. Concluding Remarks}

In this study, we considered a multi-term urn process that has a correlation in the same term and temporal correlation.
Each term is the P\'{o}lya urn model, which represents the 
correlation in the same time.
The temporal correlation represents the 
 correlation effects from the previous terms.
When the number of red balls is much smaller, 
we can obtain the Poisson process 
with the gamma distribution intensity function, the NBD process.
We introduced the temporal correlation 
as the conditional distribution for the intensity function.
This is equivalent to a self-exciting negative binomial distribution (SE-NBD) with conditional parameters.
 We referred to this process as the discrete SE-NBD process.
 This process becomes a discrete Hawkes process without correlation in the same term but with the temporal correlation between two different terms.
In the standard continuous limit of the discrete SE-NBD process, we obtained the Hawkes process.
On the other hand, taking the double-scaling limit enabled us to obtain the continuous SE-NBD process.
The difference between the continuous SE-NBD and Hawkes processes is the variance in the intensity function. 
In other words, at the limit where the intensity function becomes the delta function, the continuous SE-NBD process becomes the Hawkes process.
The continuous SE-NBD process is a marked point process. 

We observed a phase transition from the steady to the non-steady state, which is the same type of phase transition as that in the Hawkes process.
We can observe a difference in the distribution of the intensity function at the critical point.
The distribution functions of both models obey the power law at the 
critical point and have different indexes.
We applied the process to the default data to estimate the parameters.
According to our observation, the urn process is more effective for a default portfolio because of network effects.

\appendix
\section{Appendix A. Parameters for a urn process}
In this appendix we summarize the parameters for  a urn process.

$\theta_0$: Number of the red balls  at the 1-st term

$n_0$: Number of the total balls in the initial condition of  each term

$N$:Number of the balls taken out in each term

$\omega$:  In crease of the balls when we take put a ball. It is related the  correlation in the same term. 

$d_i$: Weight for  the red balls taken out at  the previous  $i$-th   terms (discount factor or the kernel function). It is one of the parameter for the temporal correlation.

$k_i$:  Number of the red balls  taken out in $i$-th term

$\tilde{\omega}$: Scale parameter for the initial condition in each term.  It is one of the  parameter for the temporal correlation.

\section{Appendix B. Proof of Eq.(\ref{Dec}) }
\begin{eqnarray}
& &\int_0^{\infty}
\mbox{Poisson}(k_1|\lambda) \cdot \mbox{Gamma}(\lambda |K_0,M_0/K_0) d \lambda
=
\int_0^{\infty} 
\frac{\lambda^{k_1}e^{-\lambda}}{k_1 !}\dot
\frac{\lambda^{K_0-1}}{\Gamma(K_0)(M_0/K_0)^{K_0}}e^{-\lambda K_0/M_0} d\lambda,
\nonumber \\
&=&
\frac{(M_0/K_0)^{-K_0}}{k_1 !\Gamma(K_0)}
\int_0^{\infty}
\frac{\Gamma(K_0+k_1)}{\Gamma(K_0+k_1)}\frac{(M_0/(M_0+K_0))^{K_0+k_1}}{(M_0/(M_0+K_0))^{K_0+k_1}}\lambda^{K_0+k_1-1}e^{-\lambda/(M_0/(M_0+K_0))}d\lambda
\nonumber \\
&=&
\frac{(M_0/K_0)^{-K_0}}{k_1 !\Gamma(K_0)}
 \Gamma(K_0+k_1)  (\frac{M_0}{M_0+K_0})^{K_0+k_1}
 \nonumber \\
 &=&
 \frac{\Gamma(K_0+k_1)}{k_1 !\Gamma(K_0)} (\frac{M_0}{M_0+K_0})^{k_1}(\frac{K_0}{M_0+K_0})^{K_0}=
 \mbox{NBD}(X_1=k_1|K_0,M_0/K_0),
\label{pg}
\end{eqnarray}
where we use 
the relation   at the third equal,
\begin{equation}
 \int_0^{\infty}
\frac{1}{\Gamma(K_0+k_1)}\frac{1}{(M_0/(M_0+K_0))^{K_0+k_1}}\lambda^{K_0+k_1-1}e^{-\lambda/(M_0/(M_0+K_0))}d\lambda=1,
\end{equation}
because it is the integral of $\mbox{Gamma}(\lambda |K_0+k_1,M_0/(M_0+K_0))$.

\section{Appendix  C. From discrete Hawkes process to Hawkes process}

This  is the discrete Hawkes process,
\begin{equation}
X_{t+1}\sim \mbox{Poisson}
\left(M_{t} \right),t\ge 0, 
\end{equation}
where 
\begin{equation}
M_t=M_0+M_0/L_0\sum_{s=1}^{t}X_s \hat{d}_{t+1-s},t\ge 1.  
\end{equation}

Here we use the counting process, $\tilde{X}_t=\sum_{i} X_i$.
The   limit $\Delta=N/n_0\rightarrow 0$ is set, as the continuous limit of process $\tilde{X}_t$.
The intensity function $\lambda_t$,
\begin{equation}
\lambda_t=\lim_{\Delta\rightarrow 0}\frac{E[\tilde{X}_{t+\Delta}-\tilde{X}_t|F_t]}{\Delta}=\lim_{\Delta\rightarrow 0}\frac{M_t}{\Delta}=(\theta_0+\tilde{\omega}\sum_{i<t} k_i\hat{d}_{t-i} ),
\end{equation}

We can  then obtain the Hawkes process,
\begin{equation}
\tilde{X}_{t+\Delta}-\tilde{X}_{t}\sim \mbox{Poisson}
\left(\theta_t\Delta \right),t\ge 0,
\end{equation}
where 
\begin{equation}
\theta_t=\theta_0+\tilde{\omega} \sum_{s<t} X_s\hat{d}_{t-s},t\ge 0.  
\end{equation}

\section{Appendix D. Marked Hawkes process}
Here we  consider  the conditional  probability in the condition that  the event occurs, $\rho(k)$, using Eq.(\ref{NBD}).
$\rho(k)$ is given as
\[
\rho(k)=\frac{1}{k\ln (\omega'+1)}
\left(\frac{\omega'}{\omega'+1}\right)^k,
\]
where $k=1,2,\cdots.$
Note that $\rho(k)$ is the gamma function with the shape parameter $0$
and does not depend on  time $t$.
The probability that an event occurs during the period $[t,t+dt]$
is
\[
\frac{\theta_t}{\omega'}\ln (\omega'+1)dt.
\]
Hence, the number of events, the marks,   is considered  IID random numbers.
In the limit $\omega'\to 0$, the distribution of the markes is
\[
\lim_{\omega\to 0}\rho(k)=\delta(k-1), 
\]
 where $\delta(x)$  is the delta function and the process reduces to the  Hawkes process.
Hence, 
SE-NBD is the marked Hawkes process \cite{oga}.

\end{document}